\documentclass{ws-ijmpd}
\usepackage[super,compress]{cite}
\usepackage{hyperref}
\usepackage{mathtools}

\begin{document}


%
\catchline{}{}{}{}{}
%

\title{Phantom energy dominated universe as a transient stage in $f(R)$ cosmology}

\author{P.H.R.S. Moraes}

\address{ITA - Instituto Tecnol\'ogico de Aeron\'autica - Departamento de F\'isica, 12228-900, S\~ao Jos\'e dos Campos, S\~ao Paulo, Brasil\\
Email: moraes.phrs@gmail.com}

\author{P.K. Sahoo}

\address{$^{\dagger}$Department of Mathematics, Birla Institute of Technology and Science-Pilani, \\ Hyderabad Campus, Hyderabad-500078, India\\
pksahoo@hyderabad.bits-pilani.ac.in}

\author{Barkha Taori}

\address{Department of Mathematics, Birla Institute of Technology and Science-Pilani, \\ Hyderabad Campus, Hyderabad-500078, India\\
barkha.m.taori@gmail.com}

\author{Parbati Sahoo}

\address{Department of Mathematics, Birla Institute of Technology and Science-Pilani, \\ Hyderabad Campus, Hyderabad-500078, India\\
sahooparbati1990@gmail.com}

\maketitle

\begin{history}
\end{history}

\begin{abstract}
The $f(R)$ theories of gravity are the most popular, simple and well succeeded extension of Einstein's General Relativity. They can account for some observational issues of standard cosmology with no need for evoking the dark sector of the universe. In the present article we will investigate LRS Bianchi type-I space-time in $f(R)$ gravity theory within the phantom energy dominated era. We show that in this formalism the phantom energy dominated universe is a transient stage and in the further stage of the universe dynamics, it is dominated, once again, by dark energy. Such an important feature is obtained from the model, rather than imposed to it, and may have a relation to loop quantum cosmology.
\end{abstract}



\tableofcontents

\section{Introduction}\label{sec:int}

The $f(R)$ theories of gravity \cite{sotiriou/2010,de_felice/2010} are the most popular and probably well succeeded alternative to the shortcomings and inconsistencies Einstein's General Relativity faces as the underlying gravitational theory \cite{capozziello/2011}-\cite{bertolami/2012}. In such a formalism, the Ricci scalar $R$ in the Einstein-Hilbert gravitational action is substituted by a generic or arbitrary function of $R$, named $f(R)$. S. Capozziello et al. have shown that the $f(R)=R^n$ gravity, with $n=3.5$, may represent a good candidate theory to solve dark matter problem \cite{capozziello/2007}. The inflationary paradigm was reviewed in the $f(R)$ gravity context in Reference \cite{sebastiani/2015}. Other important recent contributions to the $f(R)$ gravity literature can be appreciated in References \cite{battye/2018}-\cite{liang/2017}.

The main challenge in theoretical physics nowadays is dealing and understanding the dark energy, that is, the exotic kind of repulsive force that makes the universe to accelerate its expansion, rather than decelerate it, as intuition would indicate. The dark energy has been deeply investigated in $f(R)$ gravity as one can check References \cite{nojiri/2006}-\cite{amendola/2007b}, among many others. The rate of acceleration of the $f(R)$ gravity universe has been constrained via supernova Ia \cite{riess/1998}-\cite{ganeshalingam/2013} and cosmic microwave background temperature fluctuations \cite{giovi/2003,giovi/2005}.

The future dynamics of the universe is still an open question in cosmology. The present acceleration is described in standard $\Lambda$CDM model as due to a cosmological constant, whose equation of state (EoS) parameter reads $\omega=-1.006\pm0.045$ \cite{planck_collaboration/2016}. The universe dynamics today is thought to be dominated by dark energy ($\omega=-1$), anyhow, from the above observational estimation for $\omega$, one cannot distinguish precisely the regime we live nowadays ($\omega>,<$ or $=-1$). The future of the universe dynamics may be predicted once we understand the further evolution of $\omega$. Depending on the formalism, gravity theory and overall approach, $\omega$ can {\bf i)} be $-1$ forever, as standard predictions \cite{planck_collaboration/2016}; {\bf ii)} resume values $>-1/3$, which characterizes the dark energy as a transient phenomenon \cite{sahni/2003,srivastava/2007}; {\bf iii)} keep decreasing, which characterizes Big Rip models \cite{caldwell/2003}-\cite{nojiri/2005}; {\bf iv)} eternally oscillate between positive and negative values, characterizing bounce cosmological models \cite{battefeld/2015}-\cite{de_cesare/2016}.

In the present letter we will, in principle, be concerned with the third case above, named Big Rip models. In Big Rip models, the behaviour of the universe EoS makes the scale factor of the universe to become infinite at a finite time scale, named ``big rip singularity'', in which the phantom energy density also diverges. Such a cosmological property implies in the so-called ``Cosmic Doomsday'' \cite{caldwell/2003}. 

There were proposed several forms to avoid or evade the Cosmic Doomsday, such as Little Rip \cite{frampton/2011} and pseudo rip \cite{frampton/2012} models. It has already also been proposed the Big Trip, which is a cosmological event that may appear during the evolution of a cosmological wormhole embedded in a Friedmann-Lema\^itre-Robertson-Walker universe which approaches the big rip. The wormhole absorbs the phantom fluid so that it expands faster than the universe in such a way the wormhole throat radius diverges before the big rip is reached. In this alternative for evading the big rip singularity, the wormhole engulfs the entire universe, which eventually reappears from the other wormhole throat \cite{faraoni/2005,faraoni/2007}. Moreover, it was also shown that the consideration of quantum effects in the formalism may delay or even stop the singularity occurrence \cite{elizalde/2004,nojiri/2004}. 

In the present article, we will show that the phantom energy dominated era may be transient, that is, the universe not necessarily ends at the Big Rip. Particularly we will obtain Big Rip cosmological solutions in $f(R)$ gravity for the LRS Bianchi type I metric. On this regard, some cosmic observations of the Wilkinson Microwave Anisotropy Probe \cite{Komatsu/2011,Hinshaw/2013} and Planck satellites \cite{AdeA20/2016, AdeA16/2016} are detecting anisotropy in the cosmic microwave background, which are consistent with some theoretical arguments \cite{Misner/1968} and predictions of standard inflationary model \cite{Komatsu/2011}-\cite{AdeA16/2016}. As a result, it can be accepted that the early universe should be slightly anisotropic. In order to develop this scenario, one can replace some isotropic cosmological models by anisotropic inflation ones \cite{Buchert/2016}. This could be possible by employing, in cosmological models, Bianchi space-time metrics, which can be both inhomogeneous and anisotropic. In particular, Bianchi type I (hereafter BI) geometry (flat, homogeneous and anisotropic) represents the simplest extension of the standard Friedmann-Lema\^itre-Robertson-Walker metric with directional dependent scale factors. BI model provides a simplest anisotropic cosmological model that could explain the anisotropies and anomalies in the cosmic microwave background. It is worth to remark that at a special case, near the singularity, the BI metric is similar to the Kasner metric \cite{Kasner/1921}, which describes an anisotropic universe without matter.

\section{A brief summary on the Big Rip model features}\label{sec:br}
 
According to Planck satellite observations, the present value of $\omega$ is $\sim-1$ \cite{planck_collaboration/2016}. One might wonder what would happen to the universe dynamics if $\omega$ keeps decreasing, that is, if we enter a $\omega<-1$ regime. This dynamical feature can be screened in a phantom scalar field with negative kinetic term, which appears oftenly in supergravity \cite{nilles/1984} and string theory \cite{arefeva/2006}.

The fate of such a universe is quite impressive. The current abundance of dark energy in bound objects is too small to have an effect on the internal (solar systems, galaxies and clusters of galaxies) dynamics. However, for $\omega<-1$ the universe becomes increasingly dark energy-dominated, thereby the so-called phantom energy exerts growing influence on the internal dynamics and eventually the scale factor attains infinite. Ultimately, the repulsive phantom energy overcomes the forces holding the objects together and rips them apart. Not even the atoms, nuclei and nucleons would avoid getting dissociated \cite{caldwell/2003}.  
 
The condition $\omega<-1$ is not sufficient for a singularity to occur. As it was cited above, there are some proposals to prevent a Cosmic Doomsday in this scenario \cite{frampton/2011}-\cite{nojiri/2004} (check also \cite{de_haro/2012}). In the present letter, we will show that a BI metric in $f(R)$ gravity can naturally evade the end of the universe in a Big Rip, rather by further transiting it to a $\omega>-1$ era.

In the next section, we will present the main features of the $f(R)$ theory, which will be the underlying gravity theory for our present model.

\section{Structure and Mathematical Formulation of the $f(R)$ gravity}

The standard gravitational lagrangian makes use of the first order invariant Ricci scalar. Replacing this Lagrangian
by a generic function of $R$, i.e. $f(R)$, implies the modified Einstein-Hilbert action for $f(R)$ gravity as \cite{sotiriou/2010,de_felice/2010}

\begin{equation}\label{e1}
S=\int \sqrt{-g}d^{4}x\left[\frac{f(R)}{16\pi G}+L_{m}\right],
\end{equation}
with $g$ being the determinant of the metric $g_{ij}$ and from the matter Lagrangian $L_{m}$, $T_{ij}$ is the matter energy-momentum tensor defined as

\begin{equation}\label{e2}
T_{ij}=g_{ij}L_{m}-\frac{\partial L_{m}}{\partial g^{ij}}.
\end{equation}

The variation of action (\ref{e1}) with respect to the metric yields the $f(R)$ gravity field equations as 

\begin{equation}\label{e3}
F(R)R_{ij}-\frac{1}{2}f(R)g_{ij}- \nabla_{i} \nabla_{j}F(R) + g_{ij}\square F(R)= 8\pi GT_{ij}.
\end{equation}
Here, we have defined the function $F(R)$ as the derivative of $f(R)$ with respect to $R$, i.e.,   
$F(R)=\frac{\partial f(R)}{\partial R}$. Also, $\Box \equiv \nabla ^{i}\nabla _{i}$, where
$\nabla _{i}$ is the covariant derivative.

Rearranging Eq.(\ref{e3}), the field equations of $f(R)$ gravity assume the elegant form

\begin{eqnarray}\label{e4}
G_{ij}&=&\frac{1}{F(R)}{\biggl[}\frac{f(R)-RF(R)}{2}g_{ij} + \nabla_{i} \nabla_{j}F(R)\nonumber\\
&-&g_{ij}\square F(R)+8\pi GT_{ij}{\biggr]}.
\end{eqnarray}

\section{Cosmological formulation of Bianchi I phantom energy universe in $f(R)$ gravity}\label{sec:fle}

The issue of global anisotropy of our universe \cite{sung/2011} can be studied with a simple modification of the Friedmann-Lema\^itre-Robertson-Walker model. Bianchi-type models provide a generic description of spatially homogeneous anisotropic cosmologies. The simplest spatially homogeneous and anisotropic flat universe is the BI universe. The plane symmetric or LRS BI model has been proposed in order to address the issues related to the smallness in the angular power spectrum of the temperature anisotropy \cite{Campanelli/2006}-\cite{Gruppo/2007}. Some anisotropic cosmological models for BI universe as an alternative to the dark energy problem have been proposed in \cite{Rodrigues/2008}-\cite{Koivisto/2008}. In particular, the LRS BI model has been studied for inflation \cite{Koivisto/2008} and in Brans-Dicke theory \cite{Sharif/2012}. It was also analysed in $f(R,T)$ gravity models \cite{Sahoo/2015,Shamir/2015}, for which $T$ is the trace of the energy-momentum tensor. 

Here, we will consider a plane symmetric LRS BI metric of the form

\begin{equation}\label{e5}
ds^{2}=dt^2-[A^2(t)(dx^2+dy^2) + B^2(t)dz^2],
\end{equation}
where $A$ and $B$ are directional scale factors. 

The corresponding Ricci scalar reads
\begin{equation}\label{e6}
R= -2{\biggl(}2\frac{\ddot{A}}{A} + \frac{{\dot{A}}^2}{A^2} + \frac{\ddot{B}}{B} + 2\frac{\dot{A}\dot{B}}{AB}{\biggr)}.
\end{equation}

We will assume the universe is filled with a perfect fluid so that \eqref{e2} reads

\begin{equation}\label{e7}
T_{ij}=(\rho+p)u_iu_j-pg_{ij},
\end{equation}
where $\rho$ and $p$ are, respectively, the energy density and pressure of the perfect fluid and $u_i$ is
the four velocity vector, satisfying $u_iu^j =1$.

The set of field equations \eqref{e4} for metric (\ref{e5}) are obtained as:

\begin{eqnarray}
\dot{H_1}+H_1^2+\dot{H_2}+H_2^2+H_1H_2&=&\frac{1}{F(R)}{\biggl[}\frac{f(R)-RF(R)}{2}-\ddot{F}\nonumber\\&-&\dot{F}(2H_1+H_2)- 8\pi G p{\biggr]},
\end{eqnarray} \label{e8.1}

\begin{multline} \label{e8.2}
2\dot{H_1} +3H_1^2 = \frac{1}{F(R)}{\biggl[}\frac{f(R)-RF(R)}{2}-\ddot{F}-\dot{F}(2H_1+H_2) - 8\pi G p{\biggr]},
\end{multline}
\begin{multline} \label{e8.3}
H_1^2 +2H_1H_2 = \frac{1}{F(R)}{\biggl[}\frac{f(R)-RF(R)}{2}-\dot{F}(2H_1+H_2) + 8\pi G \rho{\biggr]},
\end{multline}
where dots represent differentiation with respect to time $t$. Moreover, $H_1=\frac{\dot{A}}{A}$ is the directional Hubble parameter in the directions of the $x$- and $y$-axis and $H_2=\frac{\dot{B}}{B}$ is the directional Hubble parameter in the direction of the $z$-axis.

By solving Equations (\ref{e8.2}) and (\ref{e8.3}) we obtain the following solutions for the directional scale factors $A$ and $B$:

\begin{align}
\begin{split}\label{eqn:eqlabel}
A &= ac_2^{1/3}exp{\biggl(}{\frac{c_1}{3}\int\frac{dt}{a^3}}{\biggr)},\\
B &= ac_2^{-2/3}exp{\biggl(}{\frac{-2c_1}{3}\int\frac{dt}{a^3}}{\biggr)},
\end{split}
\end{align}
where $c_1$ and $c_2$ are integrating constants and $a$ represents the average or mean scale factor such that the spatial volume is defined as $V = A^2B = a^3$. 

Moreover, the mean Hubble parameter is defined as follows:
 
\begin{equation}\label{e10}
H=\frac{1 \dot{V}}{3 V}=\frac{1}{3}(2H_1+H_2)=\frac{\dot{a}}{a}.
\end{equation}

To proceed in finding the analytical solutions we will follow Reference \cite{Nojiri/2008} in which it was considered that the mean Hubble parameter is written as

\begin{align}\label{eq11}
H=\frac{h_0}{t_0-t},
\end{align}
where $h_0$ and $t_0$ are positive constants and $H$ diverges at $t=t_0$. In \eqref{eq11}, $t_0$ is the time when the finite future singularity appears and the range lies within $0<t<t_0$ in order to carry a positive real value for the Hubble parameter $H$. 

To invoke \eqref{eq11} is to depict our model as a Big Rip model one, however as we are going to show below, by specifying a functional form for the function $f(R)$, the approach states that the universe does not end at the Big Rip. Rather, it will return to a dark energy dominated era, which apparently is the ``ground state'' or, depending on the approach assumed, the ``true vacuum'' of the universe \cite{coleman/1977}-\cite{cmr/2015}. In the following section we will investigate the main consequences and features of this scenario, starting from the epoch in which the universe dynamics is firstly dominated by dark energy. 
 
Equation (\ref{eq11}) gives 

\begin{align}\label{eq12}
 a=(t_0-t)^{-h_0},
\end{align}
\begin{align}\label{eq13}
\begin{split}
A &= \frac{c_2^{1/3}}{t_s^{h_0}} exp{\biggl[}{\frac{-c_1}{3}\frac{t_s^{3h_0 + 1}}{(3h_0 + 1)}}{\biggr]},\\
B &= \frac{c_2^{-2/3}}{t_s^{h_0}} exp{\biggl[}{\frac{2c_1}{3}\frac{t_s^{3h_0 + 1}}{(3h_0 + 1)}}{\biggr]},
\end{split}
\end{align}
where $t_s\equiv t_0-t$.

It can be seen from Equation (\ref{eq12}) that the average scale factor also diverges at $t=t_0$, which yields the Type-I curvature singularity. Moreover, it becomes a positive increasing function in the periodic range $0<t<t_0$. The deceleration parameter is obtained as $q=-\ddot{a}a/\dot{a}^2=-1-\frac{1}{h_0}$. It can be observed that for any positive value of $h_0$, $q<-1$ which describes a super exponential expansion of the universe. This expansion leads to the occurrence of a Big Rip at a finite time in the future.

Moreover, by substituting the above equations in (\ref{e6}), we obtain the curvature scalar as:
\begin{align}\label{eq14}
R= -2{\biggl[} \frac{c_1^2 t_s^{6h_0+2} + 9(2h_0^2 + h_0)}{t_s^2}{\biggr]}.
\end{align}

\section{Solutions for $f(R)$ Gravity Particular Case}


Let us consider in the formalism previously presented that $f(R)=R-\frac{a}{R}-bR^2$, where $a$ and $b$ are real numbers. This form for the $f(R)$ function was proposed by S. Nojiri and S.D. Odintsov as a possible unification of inflation and cosmic acceleration \cite{Nojiri/2003}. It was shown that the term with positive power of curvature supports the inflationary era while the term with negative power of curvature describes the cosmic acceleration.




Taking last section mechanism into account, when assuming Nojiri and Odintsov $f(R)$ functional form, the solutions for $\rho$ and $p$ read, respectively:

\begin{eqnarray}\label{eq24}
\rho&=&\frac{-1}{24\pi G}\Big[\frac{13a c_1^4 t_s^{12 h_0+6}+F_1 t_s^{6 h_0+4}+F_2 t_s^2}{4G_1(t)^3}+16 b c_1^4  t_s^{12 h_0}\nonumber\\&-&c_1^2 t_s^{6 h_0}-F_3 t_s^{6h_0-2}+\frac{9 h_0^2}{t_s^{2}}-\frac{F_4}{t_s^{4}}\Big], 
\end{eqnarray}
\begin{eqnarray}\label{eq25}
p&=&\frac{1}{24\pi G}\Big[\dfrac{11 c_1^6 t_s^{18 h_0+8}+c_1^2 t_s^{6 h_0+2}+F_5 t_s^{12 h_0+6}+F_6 t_s^{6 h_0+4}+F_7 t_s^2}{4G_1(t)^4}\nonumber\\&+&8 b c_1^4 t_s^{12 h_0}-c_1^2 t_s^{6 h_0}+F_8 t_s^{6 h_0-2}-\frac{3h_0(2+3h_0)}{t_s^{2}}+\frac{F_9}{ t_s^{4}}\Big],
\end{eqnarray}
where

\begin{eqnarray}\label{eq26}
G_1(t)=[c_1^2 (t_0-t)^{6 h_0+2}+9 h_0 (2 h_0+1)],\\
F_1=9a h_0 (37 h_0+25) c_1^2,\\
F_2=81 a h_0^2 (2 h_0+1) (23h_0+16),\\
F_3=36 b h_0 c_1^2 (7+19 h_0),\\
F_4=324b h_0^2 (2 h_0+1) (5 h_0+1),\\
F_5=3 h_0 (21 h_0+88) c_1^4,\\
F_6=27 h_0 (2 h_0+1) (5 h_0 (48 h_0+49)+12) c_1^2,\\
F_7=243 h_0^2 (2 h_0+1)^2 (23 h_0 (3 h_0+2)-12),\\
F_8=12 c_1^2 b h_0 (9 h_0+19),\\
F_9=108 b h_0 (2+3h_0)(2 h_0+1)(5 h_0-3).
\end{eqnarray} 

\begin{figure}[h!]
\centering
 \includegraphics[width=75mm]{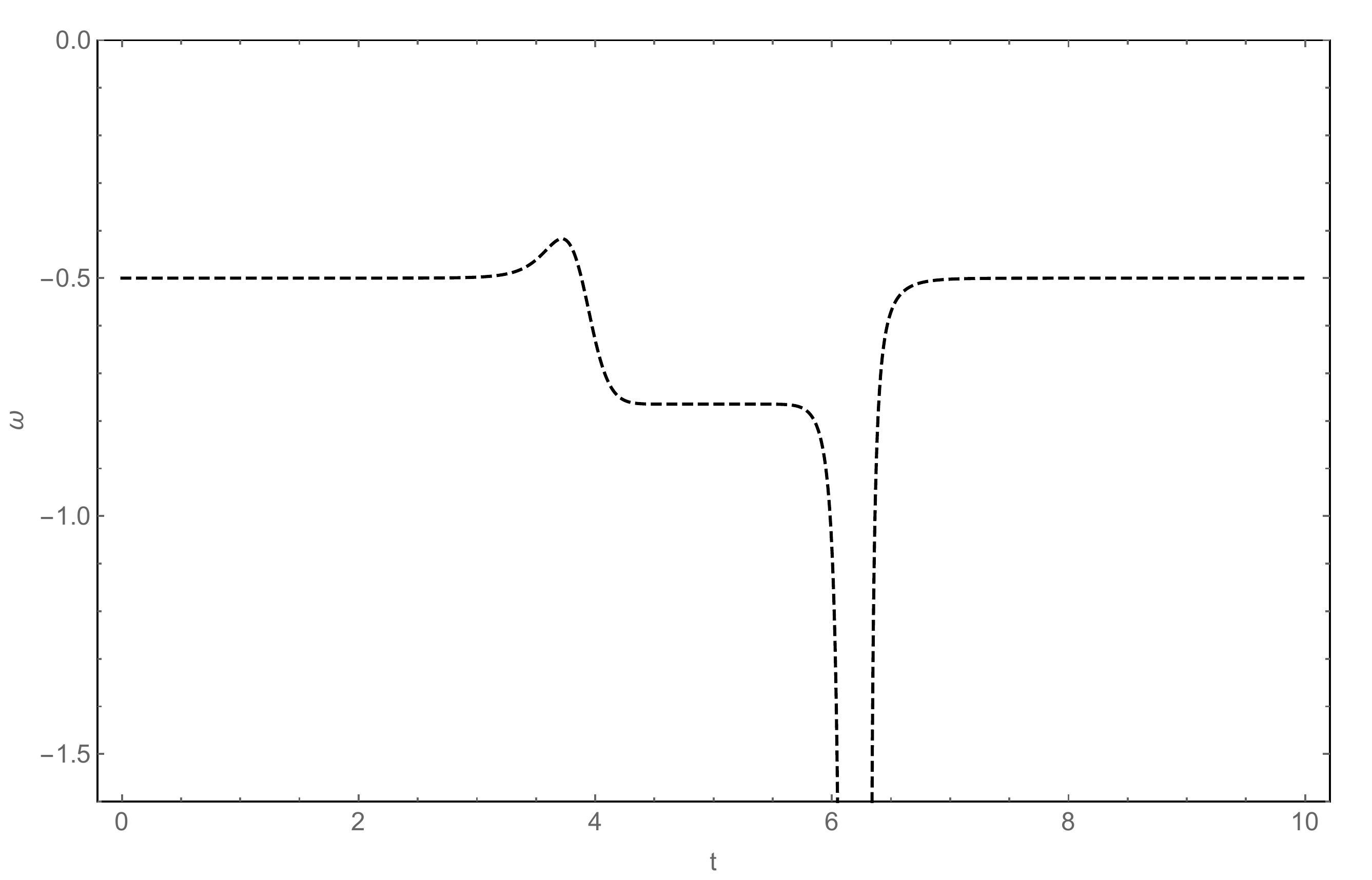}
    \caption{EoS parameter of the model for $a=7, b=-2, c_1=-2.3, h_0=1.5$ and $t_0=5$.}
\label{fig1}
\end{figure}

In Fig.\ref{fig1} below we show the time-evolution of the EoS parameter $\omega=p/\rho$ obtained from Eqs.\eqref{eq24}-\eqref{eq25}.

\section{Phantom energy dominated era as a transient stage}\label{sec:ts}

We have seen in the previous section, particularly from Fig.\ref{fig1}, that the present formalism of gravity and cosmology naturally yields to an evasion of the so-called Cosmic Doomsday. As we mentioned before, normally the evasion of the Cosmic Doomsday comes from some impositions that were not assumed in the present formalism. Rather, the universe dynamics of the present model naturally yields the phantom energy-dominated stage to be a transient phenomenon. 

From Fig.\ref{fig1} one can see that $\omega$ keeps decreasing representing what would be the Big Rip standard scenario, but surprisingly after the Rip, it increases to $>-1$ values, characterizing a second dark energy-dominated era.

As far as the present authors know, this form for evading the Cosmic Doomsday is a novelty in the literature and may be related to Loop Quantum Cosmology \cite{ashtekar/2011,bojowald/2005}, as follows.

In Loop Quantum Cosmology, the Friedmann equation reads

\begin{equation}\label{ts1}
3\left(\frac{\dot{a}}{a}\right)^2=8\pi G\rho\left(1-\frac{\rho}{\rho_c}\right),
\end{equation}
with $\rho_c\sim10^{96}$kg/m$^3$ being a critical density, for which a ``quanta'' of space cannot sustain greater values. This elegant form of writing the Friedmann equation evades the Big-Bang singularity, since $\rho_{max}=\rho_c\neq\infty$ and at $\rho_{max}$, $a=0$, that is, the scale factor also does not diverge.

By using Loop Quantum Cosmology to interpret the present scenario is to understand the later transition from a phantom energy-dominated scenario to a dark energy dominated one as a consequence of the saturation of phantom energy in the quanta of space. This saturation (attainment of $\rho_c$) would occur for the second time in the history of the universe. The first time gave rise to Big-Bang and the second one, in which $\rho$ was related to phantom energy, would lead to a later transition to a dark energy-dominated era. 

\section{Discussion}\label{sec:dis}

Big Rip models feature an intriguing scenario for the future dynamics of the universe. They have been deeply investigated in recent literature. On this regard, one can also check \cite{bouhmadi-lopez/2018}-\cite{bouhmadi-lopez/2015}. Their theoretical description is motivated by the fact that we, indeed, cannot guarantee that $\omega$ is constant today, rather, it could be still decreasing its values ($\omega\rightarrow<-1$).

The Cosmic Doomsday is the drastic universe ending in Big Rip models, in which not even the atoms, nucleons and electrons could sustain the amount of phantom energy density and would tear apart. Some attempts to evade such a catastrophic phenomenon have been proposed as one can check References \cite{frampton/2011}-\cite{nojiri/2004}.

Here in this article, a novel form for evading the Cosmic Doomsday has appeared. Rather than imposing this to the model, the universe set up yielded the universe to suffer a later transition to a dark energy-dominated era. As discussed above, this later transition could be related to Loop Quantum Cosmology, as it could be a hint that $\rho\sim\rho_c$. In this way, the Cosmic Doomsday is evaded and the universe keeps existing even after a phantom energy era.

With the purpose of being in touch with current $f(R)$ gravity and loop quantum cosmology literature, let us point some important references on such regards. In \cite{capozziello/2019} it was shown that the extra degrees of freedom of $f(R)$ gravity when compared to General Relativity can be dealt as a perfect fluid. This is a milestone of the theory and strengthens the formalism. In \cite{capozziello/2006} the possibility of having a matter dominated phase followed by an accelerated expansion in $f(R)$ cosmology formalism was discussed. The investigated cases are compatible with observational data on the Hubble parameter, which points to the reliability of $f(R)$ models. In \cite{kleidis/2018}, by using loop quantum cosmology, it was shown that during the dark energy era, a transition from a non-phantom to a phantom dark energy dominated era occurs. Furthermore, loop quantum cosmology-corrected Gauss-Bonnet $f(\mathcal{G})$ gravity is reported in \cite{kleidis/2018b}, in which $\mathcal{G}$ stands for the Gauss-Bonnet scalar.

\

{\bf Acknowledgements}

PHRSM thanks S\~ao Paulo Research Foundation (FAPESP), grant 2015/08476-0, for financial support. PKS, BT and PS acknowledges DST, New Delhi, India for providing facilities through DST-FIST lab, Department of Mathematics, where a part of this work was done.



\end{document}